\documentclass[apj]{emulateapj}

\usepackage{apjfonts}

\shorttitle{PSF subtraction algorithm for high-contrast imaging}
\shortauthors{Lafreni\`{e}re et al.}

\newcommand{\drm}{\mathrm{d}}
\newcommand{\dr}{\mathrm{d}r}
\newcommand{\dtheta}{\mathrm{d}\theta}

\begin{document}

\title{A new algorithm for point spread function subtraction in high-contrast imaging: a demonstration with angular differential imaging \footnote[1]{\lowercase{\uppercase{B}ased on observations obtained at the \uppercase{G}emini \uppercase{O}bservatory, which is operated by the \uppercase{A}ssociation of \uppercase{U}niversities for \uppercase{R}esearch in \uppercase{A}stronomy, \uppercase{I}nc., under a cooperative agreement with the \uppercase{NSF} on behalf of the \uppercase{G}emini partnership: the \uppercase{N}ational \uppercase{S}cience \uppercase{F}oundation (\uppercase{U}nited \uppercase{S}tates), the \uppercase{P}article \uppercase{P}hysics and \uppercase{A}stronomy \uppercase{R}esearch \uppercase{C}ouncil (\uppercase{U}nited \uppercase{K}ingdom), the \uppercase{N}ational \uppercase{R}esearch \uppercase{C}ouncil (\uppercase{C}anada), \uppercase{CONICYT} (\uppercase{C}hile), the \uppercase{A}ustralian \uppercase{R}esearch \uppercase{C}ouncil (\uppercase{A}ustralia), \uppercase{CNP}q (\uppercase{B}razil) and \uppercase{CONICET} (\uppercase{A}rgentina).}}}

\author{David Lafreni\`ere\altaffilmark{2}, Christian Marois\altaffilmark{2,3}, Ren\'e Doyon\altaffilmark{2}, Daniel Nadeau\altaffilmark{2}  and \'Etienne Artigau\altaffilmark{2,4}}
\altaffiltext{2}{D\'epartement de physique and Observatoire du Mont M\'egantic, Universit\'e de Montr\'eal, C.P. 6128, Succ. Centre-Ville, Montr\'eal, QC, Canada H3C 3J7}
\altaffiltext{3}{Institute of Geophysics and Planetary Physics L-413, Lawrence Livermore National Laboratory, 7000 East Ave, Livermore, CA 94550}
\altaffiltext{4}{Gemini Observatory, Southern Operations Center, Association of Universities for Research in Astronomy, Inc., Casilla 603, La Serena, Chile}
\email{david@astro.umontreal.ca cmarois@igpp.ucllnl.org \\ doyon@astro.umontreal.ca nadeau@astro.umontreal.ca \\ eartigau@gemini.edu} 

\begin{abstract} 

Direct imaging of exoplanets is limited by bright quasi-static speckles in the point spread function (PSF) of the central star. This limitation can be reduced by subtraction of reference PSF images. We have developed an algorithm to construct an optimized reference PSF image from a set of reference images. This image is built as a linear combination of the reference images available and the coefficients of the combination are optimized inside multiple subsections of the image independently to minimize the residual noise within each subsection. The algorithm developed can be used with many high-contrast imaging observing strategies relying on PSF subtraction, such as angular differential imaging (ADI), roll subtraction, spectral differential imaging, reference star observations, etc. The performance of the algorithm is demonstrated for ADI data. It is shown that for this type of data the new algorithm provides a gain in sensitivity by up to a factor 3 at small separation over the algorithm used in \citet{maroisADI}. 

\end{abstract} 

\section{Introduction}

Direct imaging of exoplanets, circumstellar disks, jets, winds or other structures around stars is difficult due to the angular proximity of the star and the very large luminosity ratios involved. Current attempts, both from the ground with adaptive optics (AO) and from space, are limited by a swarm of bright quasi-static speckles that completely mask out the faint planets or structures that are sought after \citep{schneider_roll, biller04, trident, masciadri05}. These speckles are caused mainly by imperfections in the optics and are long-lived, hence the ``quasi-static'' appellation. As the exposure time is increased, the quasi-static speckles add coherently and their intensity eventually becomes dominant over signals that add incoherently, such as sky or read noise and general (non-static) speckles.

This problem is more important closer to the star, as the relative speckle intensity is higher there, and the size of the region in which the noise is dominated by quasi-static speckles will depend upon the exposure time, the sky and read noise levels, the telescope and camera used, the target brightness, etc. For example, the observations of \citet{masciadri05} obtained at the Very Large Telescope are limited by speckles only at subarcsecond separations, while in the search for planets on wide orbits that we are currently carrying at the Gemini telescope (D. Lafreni\`ere et al., in preparation), the observations, which use longer individual exposure times (30 s), are typically limited by quasi-static speckle noise out to separations of $\sim$5\arcsec-10\arcsec. Quasi-static speckles even dominate the noise at separations well past 10\arcsec\ in the observations of the bright star Vega obtained by \citet{maroisADI} at the Gemini telescope; this also appears to be the case for similar observations obtained on the Keck and Palomar 5-m telescopes \citep{macintosh03,metchev03}. At a given angular separation, no gain in contrast is achieved by increasing the exposure time once the noise is dominated by quasi-static speckles.

When this regime is reached, it is possible to subtract the quasi-static speckles by using reference point spread function (PSF) images. A reference PSF image is any image whose subtraction from the target image would reduce the signal from the speckles while preserving that of the object sought after. For example, reference PSF images can be obtained from observations of reference stars, or from observations of the target itself obtained at different field of view orientations (e.g. \citealp{schneider_roll}), wavelengths (e.g. \citealp{racine99}), or polarizations (e.g. \citealp{kuhn01}).

Obtaining a reference PSF image highly correlated with the target image is a difficult task because even though quasi-static speckles are long lived, they still vary with time due to temperature or pressure changes, mechanical flexures, guiding errors or other phenomena \citep{trident, maroisADI}. On the other hand, even when a reference PSF image is acquired simultaneously with the target image at other wavelengths or polarizations, differential aberrations within the camera decorrelate the PSFs \citep{trident, sdi}. Thus, when trying to subtract speckles one must always work with slightly decorrelated reference PSF images and the specific way in which the available data are used to perform the subtraction may have a significant impact on the speckle noise attenuation achieved. This paper presents a way of combining reference PSF images to optimize the noise attenuation. In particular the algorithm is applied to angular differential imaging (ADI) \citep{maroisADI}, which is currently one of the most efficient quasi-static speckle suppression technique for ground-based observations. Although emphasis is given to point source detection throughout the paper, the algorithm can be optimized to search for any other structure in the close vicinity of a star.

The new reference PSF construction algorithm is presented in \S\ref{sect:alg}. Then, a review of ADI and the algorithm used by \citet{maroisADI} is presented in \S\ref{sect:adi}. In \S\ref{sect:newadi}, the new algorithm is applied to ADI and its performance is presented. The possibility of using this algorithm with other observing strategies is finally discussed in \S\ref{sect:conclusion}.

\section{Reference PSF construction by locally optimized combination of images}\label{sect:alg}

Consider a single target image, from which speckles are to be subtracted, and suppose that $N$ reference PSF images are available for this purpose. The heart of the algorithm described here is to divide the target image into subsections and to obtain, for each subsection independently, a linear combination of the reference images whose subtraction from the target image will minimize the noise. By optimizing the weights given to the $N$ available reference PSF images according to the residual noise obtained, this approach produces a representation of the target PSF image that is better than any predefined combination of the reference PSF images. Further, it is advantageous to optimize the coefficients of the linear combination for subsections of the image because the correlation between the target and the reference PSF images generally varies with position within the target image. We refer to the algorithm described here as ``locally optimized combination of images'', or LOCI.

The coefficients used for subtraction of the speckles within subsection $S^T$ of the target image are determined by a minimization of the noise within a generally larger, so-called optimization, subsection $O^T$, which encompasses $S^T$. The corresponding optimization subsections in the reference PSF images are denoted $O^n$, $n=1,\ldots,N$. 

Ideally, to achieve the optimal noise attenuation everywhere in the target image, one would want to optimize the coefficients for subsections $S^T$ that are as small as possible, ultimately consisting of a single pixel. In practice, to avoid a computationally prohibitive repetition of the algorithm, one uses subsections that contain many pixels, within which the same linear combination of reference images is used. 

While the size of the subsection $S^T$ is limited by computation resources, the size of $O^T$ is determined by the need to preserve the signal from any point source sought after. From the point of view of the algorithm described below, a point source in $O^T$ is a residual that it tries to minimize and will partially subtract. The amount of partial subtraction depends upon the fractional area of $O^T$ that is occupied by the point source. So, even though smaller optimization subsections lead to a better noise attenuation, they also lead to a larger subtraction of the signal of the point sources sought after. Thus the size of $O^T$ must be properly determined and the amount of partial subtraction of point sources must be well characterized. The area $A$ of the optimization subsection is determined by the parameter $N_A$ through the expression 

\begin{equation}\label{eq:area}
A=N_A \, \pi \left(\frac{W}{2}\right)^2
\end{equation}

\noindent where $W$ is the full-width-at-half-maximum (FWHM) of the PSF; $N_A$ thus corresponds to the number of ``PSF cores'' that fit in the optimization subsection.

If the set of reference PSF images contains target images, it is necessary to construct the optimized PSF to be subtracted from a given subsection $S^T$ by using only the subset of these images in which a companion point source appearing in $S^T$ would be displaced by at least a distance $\delta_{\rm min}$ or would have an intensity smaller by at least a factor $\alpha$ with respect to its position or intensity in the image from which speckles are to be subtracted. In other words, this subset includes all reference PSF images of index $k \in K$, where

\begin{equation}
K=\{k \in \left[1,N\right] \: : \: \left| \mathbf{r}_k-\mathbf{r}_T \right| > \delta_{\rm min}\ \lor \ f_k/f_T < \alpha \},
\end{equation}

\noindent $\mathbf{r}_T$ being any field position in the subtraction subsection of the target image and $\mathbf{r}_k$ the corresponding position in image $k$, while $f_k/f_T$ is the intensity ratio in those images of any companion sought after. If the set of reference PSF images does not contain target images, then $K=\{1,\ldots,N\}$. The parameters $\delta_{\rm min}$ and $\alpha$, when applicable, affect both the speckle noise attenuation and the amount of partial subtraction of point sources, similarly to $N_A$. The best values to use, which depend on the type of data being analyzed and the level of correlation between the target and reference images, may be determined from a comparison of the results obtained with different values, see \S\ref{sect:reg}. For the remainder of the section, it is assumed that values for $N_A$, $\delta_{\rm min}$, and $\alpha$ have been selected by the user.

The reference PSF for the optimization subsection is then constructed according to

\begin{equation}
O^R=\sum_{k \in K} c^k O^k,
\end{equation}

\noindent where the coefficients $c^k$ are to be determined by the algorithm. They are computed by minimizing the sum of the squared residuals of the subtraction of $O^R$ from $O^T$, which is given by

\begin{equation}
\sigma^2=\sum_i m_i \left( O^T_i-O^R_i\right)^2 = \sum_i m_i \left(O^T_i - \sum_k c^k O^k_i \right)^2,
\end{equation}

\noindent where $i$ denotes a pixel in the optimization subsection and $m$ is a binary mask that may be used to ignore some pixels. The quantity to minimize is a sum and can be biased by cosmic ray hits or bad pixels if they have not been properly corrected or filtered before the algorithm is used. When bad pixels remain in the image, the bias can be completely remedied by setting the mask $m$ to zero for these pixels. Generally, the fraction of pixels affected is small and their exclusion from the computation of the residuals has practically no impact on the solution found. The minimum of $\sigma^2$ occurs when all its partial derivatives with respect to the coefficients $c^k$ are equal to zero, i.e. when

\begin{equation}
\frac{\partial \sigma^2}{\partial c^j}=\sum_i -2 m_i O^j_i \left( O^T_i-\sum_k c^k O^k_i\right)=0, \quad \forall \; j \in K.
\end{equation}

\noindent Reversing the summation order and rearranging the terms we find

\begin{equation}
\sum_k c^k \left(\sum_i m_i O^j_i O^k_i\right)=\sum_i m_i O^j_i O^T_i, \quad \forall \; j \in K.
\end{equation}

\noindent This is a simple system of linear equations of the form $\mathbf{Ax}=\mathbf{b}$ where

\begin{equation}
A_{jk} = \sum_i m_i O^j_i O^k_i, \qquad x_k = c^k, \qquad b_j = \sum_i m_i O^j_i O^T_i.
\end{equation}

Solving this system gives the coefficients $c^k$ needed to construct the optimized reference PSF image for the subsection $S^T$. By construction, assuming that all the $O^k$ are linearly independent, the matrix $\mathbf{A}$ is always invertible. Thus, the system always has a unique solution, meaning that for the given optimization subsection and set $K$ the solution found leads to an \emph{absolute} minimum of the residuals. Finally, using the set of optimized coefficients, the optimized reference PSF image subsection to be subtracted from $S^T$ is constructed as 

\begin{equation}
S^R=\sum_{k \in K} c^k S^k,
\end{equation}

\noindent where $S^k$ denotes the corresponding subtraction subsection in the reference PSF image $k$.

\section{Review of angular differential imaging}\label{sect:adi}

The ADI technique, as detailed in \citet{maroisADI}, consists in acquiring a sequence of many exposures of the target using an altitude/azimuth telescope with the instrument rotator turned off (at the Cassegrain focus) or adjusted (at the Nasmyth focus) to keep the instrument and telescope optics aligned. This is a very stable configuration and ensures a high correlation of the sequence of PSF images. This setup also causes a rotation of the field of view (FOV) during the sequence. For each target image in such a sequence, it is possible to build a reference image from other target images in which any companion would be sufficiently displaced due to FOV rotation. After subtraction of the reference image, the residual images are rotated to align their FOV and co-added. Because of the rotation, the PSF residual speckle noise is averaged incoherently, ensuring an ever improving detection limit with increasing exposure time. 

In building a reference image, a compromise has to be reached between the quasi-static speckle noise correlation, which is highest for the shortest time delays between images, as shown in Figure~2 of \citet{maroisADI}, and the need to ensure a sufficient companion displacement. The minimum time delay $\tau_{\rm min}$ between an image and the ones which can be used as references decreases as the inverse of the angular separation. Accordingly, it is possible to use images more closely separated in time to build the reference image at larger angular separations. 

In the speckle subtraction algorithm used in \citet{maroisADI} (see their \S5.2 and their Table 2), the first step is to subtract the median of all the images from each individual image. Each target image is then broken into many annuli to reflect the dependence of $\tau_{\rm min}$ on the distance from the center of the PSF. A reference image is obtained within each annulus by median combining the four images obtained closest in time but at least $\tau_{\rm min}$ from the target image. The intensity of this reference image is also scaled to minimize the noise after reference subtraction. All the resulting images are then rotated to align their FOV and a median is taken over them.

\section{Application of the LOCI algorithm to angular differential imaging data}\label{sect:newadi}

\subsection{Definition of the arbitrary parameters specific to ADI}\label{sect:reg}

As mentioned in \S\ref{sect:alg}, some parameters must be chosen by the user before the algorithm is used. For ADI data, as images are generally acquired in a single bandpass, the parameter $\alpha$ does not apply. On the other hand, the area and shape of the subtraction and optimization subsections must be defined as well as the minimum displacement $\delta_{\rm min}$ between sources in the target and reference images. 

The dependence of $\tau_{\rm min}$ on angular separation suggests the use of annular geometry for the subtraction subsections. These subsections are obtained by further dividing the annuli azimuthally to reduce their spatial extent, which enables a better fit of local PSF variations as explained above. Since $\tau_{\rm min}$ is proportional to $1/r$, the set of images that can be used to construct a reference PSF changes rapidly with radius at small separation and it is best to use narrow subsections at small radii to ensure that the largest possible set of reference PSF images is used at any separation. The subtraction subsections $S^T$ are described by their inner radius $r$, mean angular position $\phi$, radial width $\drm r$, and angular width $\Delta\phi$.

The optimization subsections $O^T$ share the same inner radius, mean angular position, and angular width as their corresponding subtraction subsection. As explained in \S\ref{sect:alg}, the area $A$ of the optimization subsections has to be chosen to maximize noise attenuation while minimizing point source subtraction. For an annular subsection $O^T$ of inner radius $r$, radial width $\Delta r$, and azimuthal width $(r+\Delta r/2)\Delta\phi$,

\begin{equation}
A=\Delta r (r+\Delta r/2)\Delta\phi.
\end{equation}

We define the ratio of the radial and azimuthal widths of the optimization subsections as

\begin{equation}
g= \frac{\Delta r}{(r+\Delta r/2)\Delta \phi} =\Delta r^2/A.
\end{equation}

\noindent Then, by Eq.(\ref{eq:area}), $\Delta r$ and $\Delta \phi$ are uniquely determined by the parameters $g$ and $N_A$ and the PSF width $W$:

\begin{equation}
\Delta r= \sqrt{\pi g N_A W^2 /4},
\end{equation}
and
\begin{equation}
\Delta\phi= \left( \frac{g}{2}+\frac{2r}{W}\sqrt{\frac{g}{\pi N_A}}\right)^{-1}.
\end{equation}

\begin{figure}
\epsscale{1}
\plotone{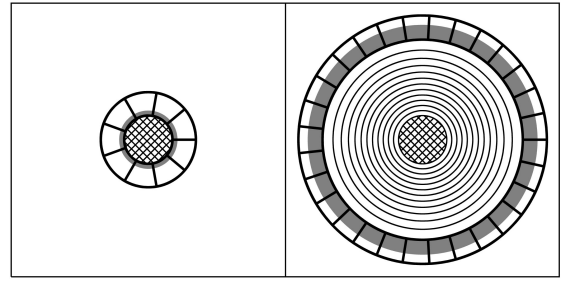}
\caption{\label{fig:regions} Example of subtraction ({\it shaded in grey}) and optimization ({\it delimited by thick lines}) subsections for ADI using the procedure of \S\ref{sect:reg}. The left and right panels show the subtraction and optimization subsections for the 1$^{\rm st}$ and 13$^{\rm th}$ subtraction annuli respectively. In the right panel, the first 12 subtraction annuli (of width $\dr$) are marked by thin lines; $\dr$ increases with separation in this specific example. The central circle (cross-hatched) represents the saturated region.}
\end{figure}

\noindent The optimization subsections were chosen not to be centered radially on the subtraction subsections but to extend to larger radii because, in the optimization, the radial dependence of the PSF noise gives more weight to the inner pixels, i.e. to the pixels in $S^T$. Figure~\ref{fig:regions} shows an example of subsections that can be used with this procedure.

For ADI data, the set of target images is the same as the set of reference PSF images. The subset of images that can be used as references for a given subsection $S^T$ of a given target image depends upon the parameter $\delta_{\rm min}$ introduced in \S\ref{sect:alg}, whose value is set by the parameter $N_\delta$ through the expression

\begin{equation}
\delta_{\rm min}=N_\delta W+r \, \dtheta_n,
\end{equation}

\noindent where $\dtheta_n$ is the angle of FOV rotation that occurred during exposure $n$. The last term of the expression above represents the azimuthal smearing of an off-axis point source that occurs during an exposure due to FOV rotation. The parameter $N_\delta$ represents the minimum gap allowed, in units of the PSF FWHM, between a source position in image $n$ and the corresponding positions in the images used as references.

The values of $N_A$, $g$, $\dr$ and $N_\delta$ that maximize the sensitivity to faint point sources will be determined in the next section using real data.

\subsection{Parameter determination}\label{sect:opt}

Observations of the star HD97334b (G0V, $H=5$) were used to determine the values of the algorithm parameters. These observations are part of the Gemini Deep Planet Survey (GDPS, D. Lafreni\`ere et al., in preparation), which is an ongoing direct imaging search for Jupiter mass planets on large orbits ($>40$~AU) around young nearby stars ($\sim$100 Myr, $<$35~pc). This particular dataset consists in a sequence of $90$ 30-s images in the CH$_4$-short ($1.58 \; \mu$m, $6.5$\%) filter obtained with ALTAIR/NIRI at the Gemini North telescope (program GN-2005A-Q-16). The $f/32$ focal ratio of the camera and 8-m primary mirror diameter lead to 0\farcs022 pixel$^{-1}$. The images are saturated inside a radius of $\sim$0\farcs7 from the PSF center. Short unsaturated exposures were acquired before and after the saturated sequence to calibrate photometry and detection limits. The corrected PSF FWHM was measured to be $3.4$ pixels, or 0\farcs074, and the Strehl ratio was $\sim$$16$\%. The Cassegrain rotator was fixed during all observations. Basic image reduction and registering was done as in \citet{maroisADI}.

\begin{figure}
\epsscale{1}
\plotone{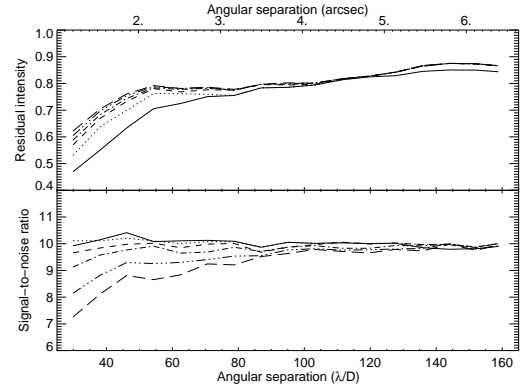}
\caption{\label{fig:nd} Average residual intensity of the artificial point sources normalized to their initial intensity (\emph{top}), and their S/N ratio (\emph{bottom}) as a function of angular separation, for different values of $N_\delta$. The solid, dotted, dashed, dot-dashed, triple-dot-dashed, and long-dashed curves are respectively for $N_\delta=0.25$, 0.5, 0.75, 1.0, 1.5, and 2.0.}
\end{figure}

\begin{figure}
\epsscale{1}
\plotone{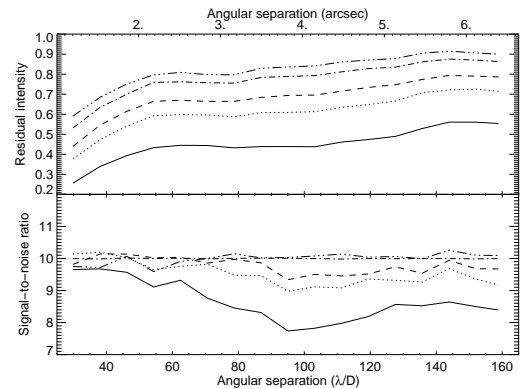}
\caption{\label{fig:na} 
Average normalized residual intensity (\emph{top}) and S/N ratio (\emph{bottom}) as a function of angular separation for different values of $N_A$. The solid, dotted, dashed, dot-dashed, and triple-dot-dashed curves are respectively for $N_A=50$, 100, 150, 300, and 500.}
\end{figure}

\begin{figure}
\epsscale{1}
\plotone{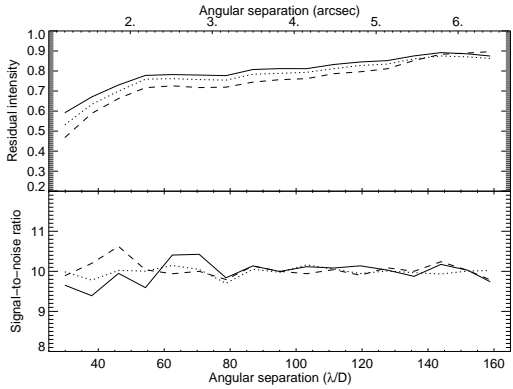}
\caption{\label{fig:geom} 
Average normalized residual intensity (\emph{top}) and S/N ratio (\emph{bottom}) as a function of angular separation for different values of $g$. The solid, dotted, and dashed curves are respectively for $g=0.5$, 1, and 2.}
\end{figure}

\begin{figure}
\epsscale{1}
\plotone{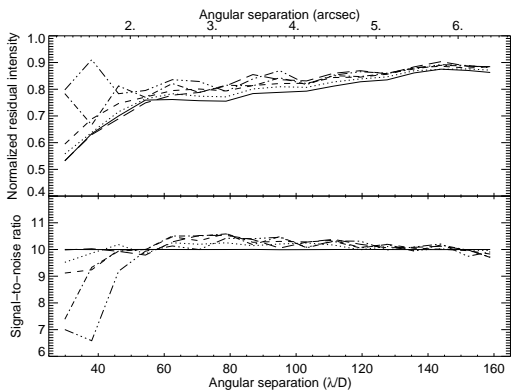}
\caption{\label{fig:dr} 
Average normalized residual intensity (\emph{top}) and S/N ratio (\emph{bottom}) as a function of angular separation for different values of $\dr$. The solid, dotted, dashed, dot-dashed, triple-dot-dashed, and long-dashed curves are respectively for $\dr=1.5$, 3, 6, 9, 15, and $\dr$ varying with radius (see text).}
\end{figure}

The same procedure was used for optimizing each of $N_\delta$, $N_A$, $g$ and $\dr$. First, the unsaturated PSF image was used to introduce artificial point sources to the images at angular separations in the range 50-300 pixels (27-160 $\lambda/D$) in steps of 5 pixels (2.75 $\lambda/D$). Each artificial source was smeared according to its displacement during an integration, and its intensity was set so that its S/N would be $\sim$10 in the final residual combination. Next, a symmetric radial profile was subtracted from each image to remove the seeing halo. Then the subtraction algorithm was run on the sequence of images with a range of values for the parameter under consideration. Finally, the noise and the flux of each artificial point source in an aperture diameter of one FWHM were measured in the residual image. This process was repeated $50$ times by placing the artificial sources at different angular positions each time. The trial values for the optimization of each parameter are listed in Table~\ref{tab:optval}. For $\drm r$ either a fixed value is used throughout the image or one that varies from 1.5 to 15 in units of the PSF FWHM. The optimal value of a parameter was determined recursively, with the values of the other parameters set first to the medians of the values listed in Table~\ref{tab:optval} and then to their most recently determined optimal value except for $\dr$ set at a fixed value of 1.5. The results are shown in Figures~\ref{fig:nd}-\ref{fig:dr}.

\begin{deluxetable}{ccc}
\tablewidth{0pt}
\tablecolumns{3}
\tablecaption{\label{tab:optval} Parameter values used for optimization}
\tablehead{ \colhead{Parameter} & \colhead{Trial values} & \colhead{Adopted value} }
\startdata
$N_\delta$ & 0.25, 0.5, 0.75, 1.0, 1.5, 2.0 & 0.5 \\
$N_A$      & 50, 100, 150, 300, 500         & 300 \\
$g$        & 0.5, 1.0, 2.0                  & 1.0 \\
$\dr$      & 1.5, 3, 6, 9, 15, (1.5-15)\tablenotemark{a} & (1.5-15)\tablenotemark{a} \\
\enddata
\tablenotetext{a}{$\dr$ equal to 1.5 for separations less than 60 $\lambda/D$ and 15 for larger separations.}
\end{deluxetable}

As can be seen in Figure~\ref{fig:nd}, the minimum spacing has little impact on the recovered flux at separations $\gtrsim 100 \ \lambda/D$, where $\sim$80-90\% of the flux is recovered independently of $N_\delta$. However, at small separations the effect is important and significant loss in signal occurs, particularly for the smallest minimum displacements. This is because the fraction of images in the subset $K$ for which the point source partially overlaps that in the target image is greater for smaller separations, where linear motion of the point source is slower. The best overall S/N is obtained with $N_\delta=0.5$, for which the loss in the recovered flux is more than compensated by the improvement in quasi-static speckle noise attenuation.

Figure~\ref{fig:na} shows that the residual signal of point sources is strongly dependent upon the size of the optimization subsections, as expected from the discussion in \S\ref{sect:alg}. When $N_A$ is too small, the gain in attenuation is not sufficient to compensate for the larger point source subtraction and lower S/N ratios are obtained, especially at large separations. On the other hand, when $N_A$ is too large, the quasi-static speckles are not subtracted as efficiently at small separations and lower S/N ratios result. A value of $N_A=300$ provides the best overall S/N ratio.

The parameter $g$ has little effect on the performance, see Figure~\ref{fig:geom}, although for angular separations $\lesssim50\ \lambda/D$ regions more extended radially ($g=2$) fare slightly better than regions more extended azimuthally. Nevertheless, we adopt $g=1$ as the optimal value.

Finally, Figure~\ref{fig:dr} shows that at small separations ($\lesssim60\ \lambda/D$), a $\dr \gtrsim 6$ leads to a lower S/N ratio because it poorly matches the evolution of $\tau_{\rm min}$ with separation, as expected. Since a larger $\dr$ leads to a faster execution of the algorithm, because fewer subtraction subsections are required to cover the entire image, we use as the optimal value a $\dr$ equal to 1.5 for separations less than 60~$\lambda/D$ and 15 for larger separations.

The optimal parameter values may vary slightly from those found above for other sets of data depending on the telescope, instrument, seeing, FOV rotation rate, target brightness, etc. They are optimized here for a specific set of data only to illustrate the potential of the LOCI algorithm for ADI. For all computations that follow, the optimal values listed in Table~\ref{tab:optval} are used.

\subsection{Point source photometry}\label{sect:phot}

\begin{figure}
\epsscale{1}
\plotone{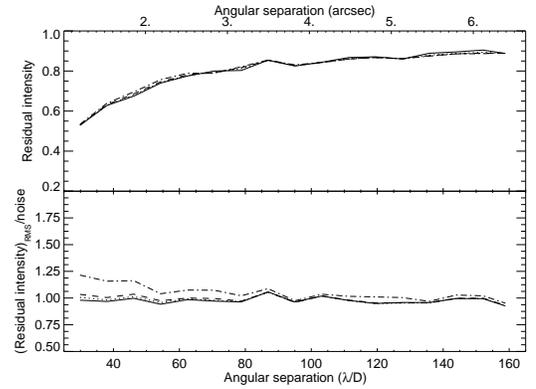}
\caption{\label{fig:phot} Average normalized residual intensity (\emph{top}) and ratio of the measured dispersion of the residual intensity of sources over the residual noise (\emph{bottom}). The solid, dotted, dashed, and dot-dashed lines are respectively for point source intensities yielding S/N of 3, 6, 10, and 25 in the final residual image.}
\end{figure}

Since the algorithm reduces the flux of point sources significantly, especially at small separations, it is important to verify that the true flux can be recovered accurately and that the uncertainty on this value can be well determined. The algorithm was run on the sequence of images, with artificial companions of various intensities added at all angular separations in the range 50-300 pixels (27-160 $\lambda/D$) by steps of 5 pixels (2.75 $\lambda/D$). Four intensities were used, yielding S/N of 3, 6, 10 and 25 in the final residual image. This process was repeated 50 times with the sources at different azimuthal positions. The mean normalized residual source intensities and residual intensity dispersions over the 50 azimuthal positions were then computed and the results are shown in Figure~\ref{fig:phot}.

The top panel of this figure shows that the normalized residual intensities do not vary with the intensity of the sources, i.e. \emph{the fraction of the signal of a source that is subtracted by the algorithm is independent of the source brightness}. Hence, a normalized residual intensity curve obtained by implanting artificial point sources of a given brightness can be used to calculate the true flux of sources of any brightness and to correct the detection limit curve computed from the variance of the residual noise.

The bottom panel of Figure~\ref{fig:phot} shows that the noise measured in the residual image is an adequate measure of the uncertainty on the intensity of sources at 10$\sigma$ or less. For brighter sources ($\sim$25$\sigma$), the uncertainty is slightly larger for small separations. This is probably due to the larger bias introduced by brighter point sources and the more important dependence of the amount of partial subtraction on the specific PSF structure underlying the point source in regions strongly dominated by quasi-static speckle noise. Thus, the noise in the residuals may be used as the uncertainty on the flux for most sources but it may be necessary to carry out an analysis using artificial point sources for brighter sources at small separations.

\subsection{Comparison with previous algorithm}

\begin{figure}
\epsscale{1}
\plotone{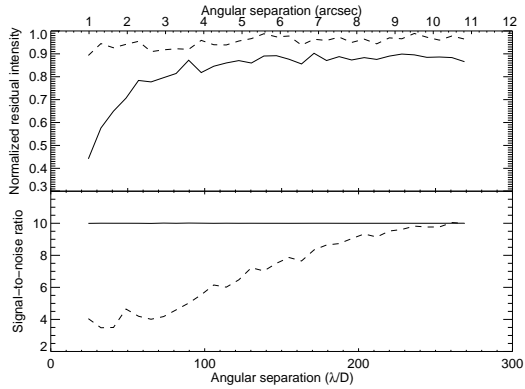}
\caption{\label{fig:comp} 
Average normalized residual intensity (\emph{top}) and S/N ratio (\emph{bottom}) as a function of angular separation for the LOCI algorithm ({\it solid line}) and the algorithm of \citet{maroisADI} ({\it dashed line}).}
\end{figure}

\begin{figure}
\epsscale{1}
\plotone{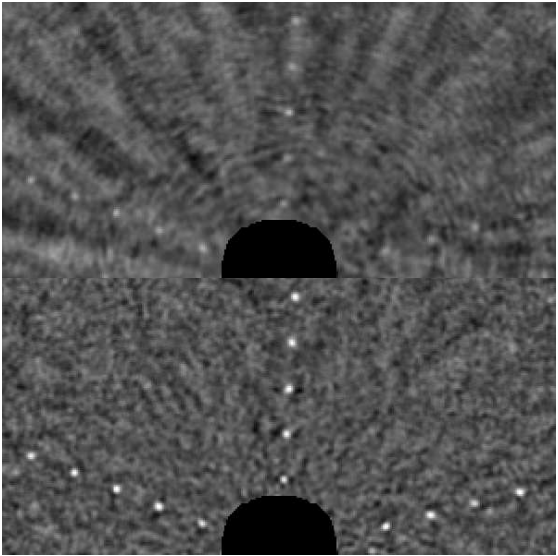}
\caption{\label{fig:im} Residual S/N image (including artificial point sources) using the algorithm of \citet{maroisADI} (\emph{top}) and the LOCI algorithm (\emph{bottom}). Both panels are shown with a (-5,+10) intensity range. Each panel is 6\farcs5 by 3\farcs25. The images have been convolved by a circular aperture of diameter equal to $W$. The saturated region at the center of the PSF is masked out.}
\end{figure}

A comparison of the LOCI algorithm with that used by \citet{maroisADI} is presented. Artificial point sources were added to the images at several separations in the range 40-500 pixels (22-275 $\lambda/D$) by steps of 5 pixels (2.75 $\lambda/D$). The intensities of the artificial sources were adjusted to yield a final S/N$\sim$10 with the LOCI algorithm. Both subtraction algorithms were then run on the images. This was repeated 25 times with the artificial sources at different azimuthal positions. The mean residual intensity and S/N over the 25 azimuthal positions were then computed for each algorithm and separation. The results are shown in Figure~\ref{fig:comp}. At all separations, the LOCI algorithm yields a S/N that is better or equal to that obtained with the algorithm of \citet{maroisADI}. The gain is highest at small separations, where it reaches a factor $\sim$3, and steadily decreases for larger separations. The decrease is most likely due to the increasing relative importance of sky background noise. A comparison of the residual image of the two algorithms is shown in Figure~\ref{fig:im}; the lower level of noise of the LOCI algorithm is clearly visible. The new algorithm yields a better attenuation because it can adapt more easily to temporal and spatial variations of the PSF quasi-static speckle pattern by using all the images available with proper weights (the coefficients) and optimizing the reference image combination in smaller subsections.

\begin{figure}
\epsscale{1}
\plotone{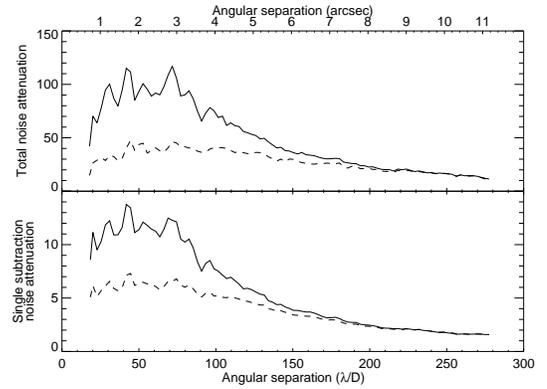}
\caption{\label{fig:att} Noise attenuation resulting from a single reference image subtraction (\emph{bottom}) and total noise attenuation (\emph{top}). The noise attenuation is defined as the ratio of the noise in the target image over that in the residual image; the noise is computed as the standard deviation of the pixel values inside an annulus of width $\sim$1 PSF FWHM. The dashed and solid lines are respectively for the algorithm of \citet{maroisADI} and the LOCI algorithm. The attenuations have been corrected for the partial subtraction of point sources. Before computation of the initial noise level, a $7\times7$ PSF FWHM median filter was subtracted from the images to remove the low spatial frequency structures that do not prevent point source detection.}
\end{figure}

\begin{figure}
\epsscale{1}
\plotone{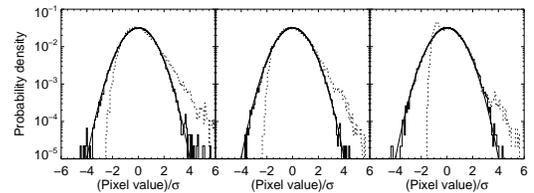}
\caption{\label{fig:hist} Statistical distributions of the pixel values of one original S/N image after subtraction of a radial profile (\emph{dotted line}) and of the final S/N residual image (\emph{solid line}) obtained with the LOCI algorithm. From left to right, the three panels are for angular separations of 25, 50 and 150 $\lambda/D$ respectively. Both images have been convolved by a circular aperture of diameter equal to $W$ and annuli of area equal to $5000\ \pi (W/2)^2$  were used to obtain the distributions at each separation. The continuous solid curve shows a Gaussian distribution of unit standard deviation.}
\end{figure}

\begin{figure}
\epsscale{1}
\plotone{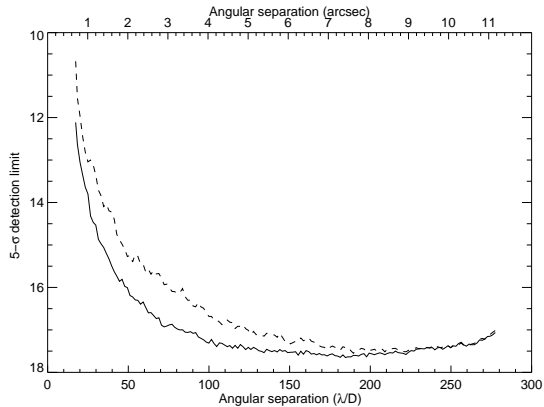}
\caption{\label{fig:limit} Point source detection limit. The dashed and solid lines are respectively for the algorithm of \citet{maroisADI} and the LOCI algorithm. The detection limits have been corrected for the partial subtraction of point sources, for the anisoplanatism observed with ALTAIR and for the slight smearing of point sources during an exposure due to FOV rotation.}
\end{figure}

The subtraction algorithms were then applied to the original sequence of images, i.e. without artificial sources, to compare the quasi-static speckle noise attenuation they provide and the detection limits they achieve. The quasi-static speckle noise attenuation is shown in Figure~\ref{fig:att}; a single subtraction using the LOCI algorithm provides an attenuation of $\sim$10-12 at separations of 1-3 arcsec. The formulation of a simple and universal criterion for speckle-limited point source detection is usually complicated because the distribution of speckle noise is non Gaussian \citep{schneider_roll, aime04, marois_these, fitzgerald06}; it possesses an important tail at the higher end. However, ADI leads to residuals whose distribution closely resembles a Gaussian; this is studied in more detail elsewhere (C. Marois et al., in preparation). This was indeed verified for the data presented here, see Figure~\ref{fig:hist}; a few events above a Gaussian distribution are seen only at the smallest angular separations. A 5$\sigma$ threshold is thus adequate for estimating detection limits. The final 5$\sigma$ detection limits in difference of magnitudes reach 13.9, 16.1 and 16.9 at angular separations of 1, 2 and 3 arcsec respectively, see Figure~\ref{fig:limit}. The speckle noise attenuation and the detection limits have been properly corrected for the partial loss of signal of point sources as measured from the residual signal of artificial sources.

Comparison of the two algorithms were made using a few different observation sequences and similar results were obtained every time.

\section{Conclusion}\label{sect:conclusion}

An algorithm to construct an optimized reference PSF image used to subtract the speckle noise and improve the sensitivity to faint companion detection was developed and tested. For a given target image limited by speckle noise, the algorithm linearly combines many reference PSF images such that the subtraction of this combination from the target image minimizes the speckle noise. Optimization of the coefficients of the linear combination is done for multiple subsections of the image independently and the procedure ensures that the minimum residual noise is reached within each subsection. The application of the algorithm to ADI yielded a factor of up to 3 improvement at small separations over the algorithm used in \citet{maroisADI}.

The algorithm presented in \S\ref{sect:alg} is general and can be used with most high contrast imaging observations aimed at finding point sources. In particular, it can be used with a sequence of images of the target itself obtained at different FOV orientations (ADI, roll subtraction for HST \citep{schneider_roll}, ground-based observations with discrete instrument rotations, etc.), with images of the same target at different wavelengths (simultaneous spectral differential imaging (SSDI, \citealp{racine99, marois2000}) or non-simultaneous spectral differential imaging (NSDI) with, for example, a tunable filter) or with images of reference stars acquired with the same instrument in a similar configuration. The latter could be particularly interesting for HST for which the PSF is more stable than at any ground-based telescope and for which suitable observations of reference stars may be readily retrieved from the archive. This should also be the case for the James Webb Space Telescope (JWST), whose temperature is expected to be much more stable as a result of its more stable environment. Future ground-based instrumentation designed specifically for finding exoplanets will have a small FOV, rendering SSDI inefficient to detect planets whose spectrum has no steep feature and ADI inefficient because of the very long time baseline required for sufficient rotation. For such cases, discrete instrument rotations may be critical and the algorithm developed here could be used directly. The Fine Guidance Sensor onboard JWST \citep{rowlands_fgs}, which will include a tunable filter imager \citep{rowlands_tfi} and coronagraph \citep{doyon_tfc}, is a very interesting prospect for NSDI. Again, the algorithm developed here could be applied directly to this case.

\acknowledgments

This work was supported in part through grants from the Natural Sciences and Engineering Research Council, Canada, from Fonds Qu\'eb\'ecois de la Recherche sur la Nature et les Technologies and from the Facult\'e des \'Etudes Sup\'erieures de l'Universit\'e de Montr\'eal. This research was performed in part under the auspices of the US Department of Energy by the University of California, Lawrence Livermore National Laboratory under contract W-7405-ENG-48, and also supported in part by the National Science Foundation Science and Technology Center for Adaptive Optics, managed by the University of California at Santa Cruz under cooperative agreement AST 98-76783.

\end{document}